\documentclass[sort&compress,5p]{elsarticle}

\usepackage{upgreek}

\begin{document}

\begin{frontmatter}
\title{The N=126 Factory: A New Multi-Nucleon Transfer Reaction Facility}
\author[ANL,TUD]{A.A. Valverde\corref{cor1}}\ead{avalverde@ikp.tu-darmstadt.de}
\author[ANL]{M.S. Martin}\ead{matthew.martin@anl.gov}
\author[ND]{W.S. Porter}\ead{wporter@nd.edu}
\author[ANL,ND]{A.M. Houff}
\author[ANL,CHI]{C.M. Baruch}
\author[ND]{M. Brodeur}
\author[ANL]{J.A. Clark}
\author[ANL]{Y. Cho}
\author[ANL]{A. Jacobs}
\author[ANL]{R.A.~Knaack}
\author[TUD]{F. K\"ohler}
\author[TUD]{K. K\"onig}
\author[ANL]{O.S. Kubiniec}
\author[ANL,CHI]{A. LaLiberte\fnref{fn2}}
\author[ANL,ND]{B. Liu\fnref{fn3}}
\author[ANL]{B. Maass}
\author[ANL,ND]{A. Mitra}
\author[ANL]{P. Mueller}
\author[ANL]{C.~M\"uller-Gatermann}
\author[TUD]{W. N\"ortersh\"auser}
\author[ANL]{M.B. Oberling}
\author[TUD]{J. Palmes}
\author[ANL,ND]{C. Quick\fnref{fn1}}
\author[ND]{E.S.C. Ribeiro}
\author[ANL]{J. Rohrer}
\author[ANL,CHI]{G.~Savard}
\author[TUD]{J. Spahn}

\address[ANL]{Physics Division, Argonne National Laboratory, Lemont, IL 60439, USA}
\address[TUD]{Institut für Kernphysik, TU Darmstadt, 64289 Darmstadt, Germany}
\address[ND]{Department of Physics \& Astronomy, University of Notre Dame, Notre Dame, IN 46556, USA}
\address[CHI]{Department of Physics, University of Chicago, Chicago, IL 60637, USA}

\cortext[cor1]{Corresponding author}

\fntext[fn2]{Present address: Department of Physics and Astronomy, Michigan State University, East Lansing, MI 48824, USA }
\fntext[fn3]{Present address: Nuclear Science Division, Lawrence Berkeley National Laboratory, Berkeley, CA  94720, USA}
\fntext[fn1]{Present address: Department  of Physics and Astronomy, University of Tennessee, Knoxville, TN 37996, USA}

\begin{abstract}
Multi-nucleon transfer (MNT) reactions between two heavy ions offer an effective method of producing heavy, neutron-rich nuclei that cannot currently be accessed efficiently using traditional production techniques. These nuclei are important for understanding many astrophysical phenomena, such as the formation of the r-process $A\sim 195$ abundance peak. The N=126 Factory currently commissioning at Argonne National Laboratory's ATLAS facility will make use of these reactions to allow for the study of these nuclei. To convert MNT reaction products, which have a wide angular distribution, into a collimated, bunched beam suitable for experiments, a series of apparatus will be used. These start with a large-volume gas catcher for stopping the reaction products, which are then extracted through a radiofrequency quadrupole ion guide, undergo preliminary dipole magnetic separation, cooling and bunching in a Cooler-Buncher, and final separation using a multi-reflection time-of-flight mass separator, before final delivery to experimental systems. 
\end{abstract}
\end{frontmatter}

\section{Introduction}
Precision nuclear data throughout the nuclear landscape, including masses, $\beta$-decay half lives, and neutron capture cross sections, are essential in order to properly model nucleosynthetic pathways and accurately predict observed elemental abundance patterns. Recent sensitivity studies \cite{Mumpower16} have highlighted the data in the neutron-rich region surrounding the $N=126$ shell closure as having high impact, at least in part due to its limited availability. Additional motivation comes from possible nucleon shell evolution or other structural effects in this region. 

Data in neutron-rich nuclei in the vicinity of the $N=126$ shell closure are limited in part due to the difficulties producing beams of these nuclei via traditional methods, such as fragmentation, fusion evaporation, spallation and fission. Multi-nucleon transfer (MNT) reactions have significantly higher production cross sections for neutron-rich $N\sim 126$ nuclei \cite{Zagrebaev08, Hirayama2016,Watanabe15}, however as these are peripheral reactions close to the grazing angle, there exist substantial experimental challenges in manipulating the MNT products into a beam which can be directed towards experimental endstations.

The N=126 Factory under development at Argonne National Laboratory's (ANL's) ATLAS facility will take advantage of technologies developed for the CARIBU \cite{Savard2008} and nuCARIBU \cite{McLain2022} facilities to catch and manipulate the MNT products into an ion beam.
\section{Facility Layout} \label{sec: facility}
\begin{figure}[b!]
\includegraphics[width=\columnwidth]{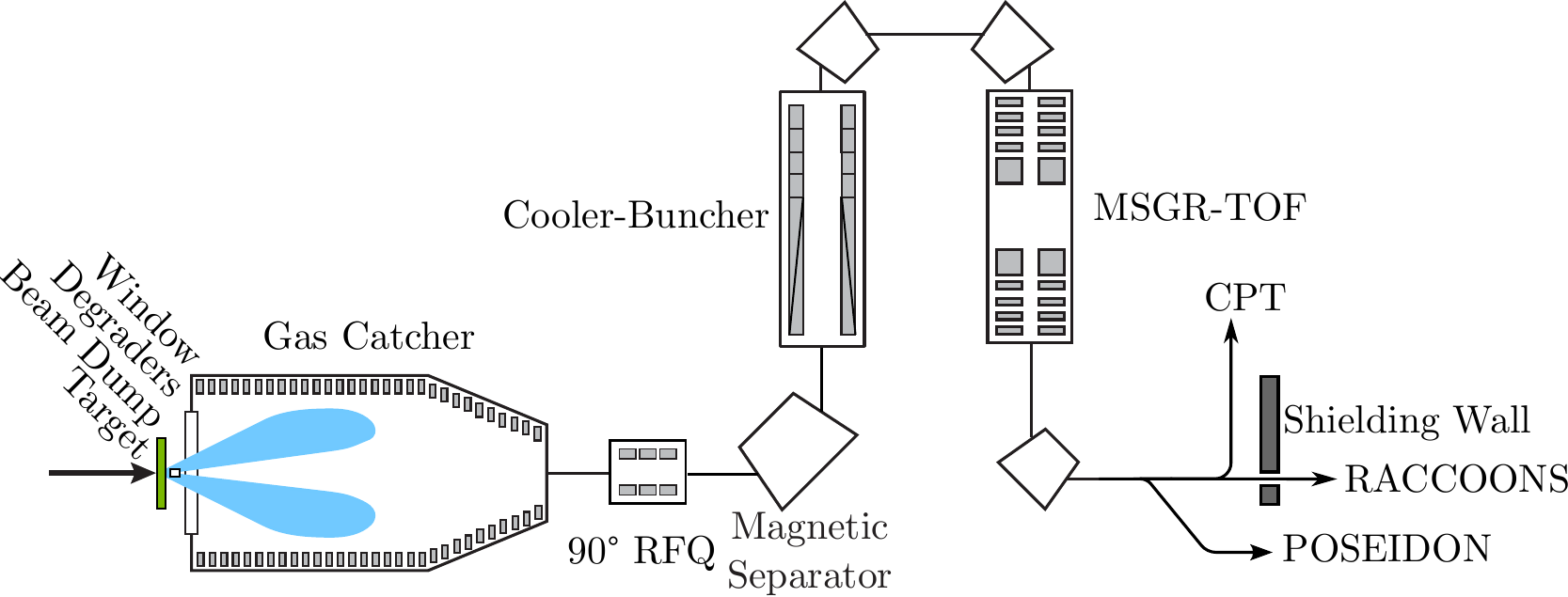}
\caption{Schematic diagram of the N=126 Factory, illustrating beam production, major components, and experimental endstations. \label{fig:Diagram}}
\end{figure}
Located at ANL's ATLAS accelerator facility, the N=126 Factory will take advantage of ATLAS's ability to deliver high-intensity stable beams to produce rare isotopes far from stability and deliver them to experimental endstations. A diagram of the layout of the facility can be seen in Fig.\,\ref{fig:Diagram}, showing schematically the important components of the facility for producing, stopping, separating, preparing, and delivering bunched, isotopically pure, low emittance beams for a low-energy experimental program.

\subsection{Target and Degraders}
\begin{figure}
\includegraphics[width=\columnwidth]{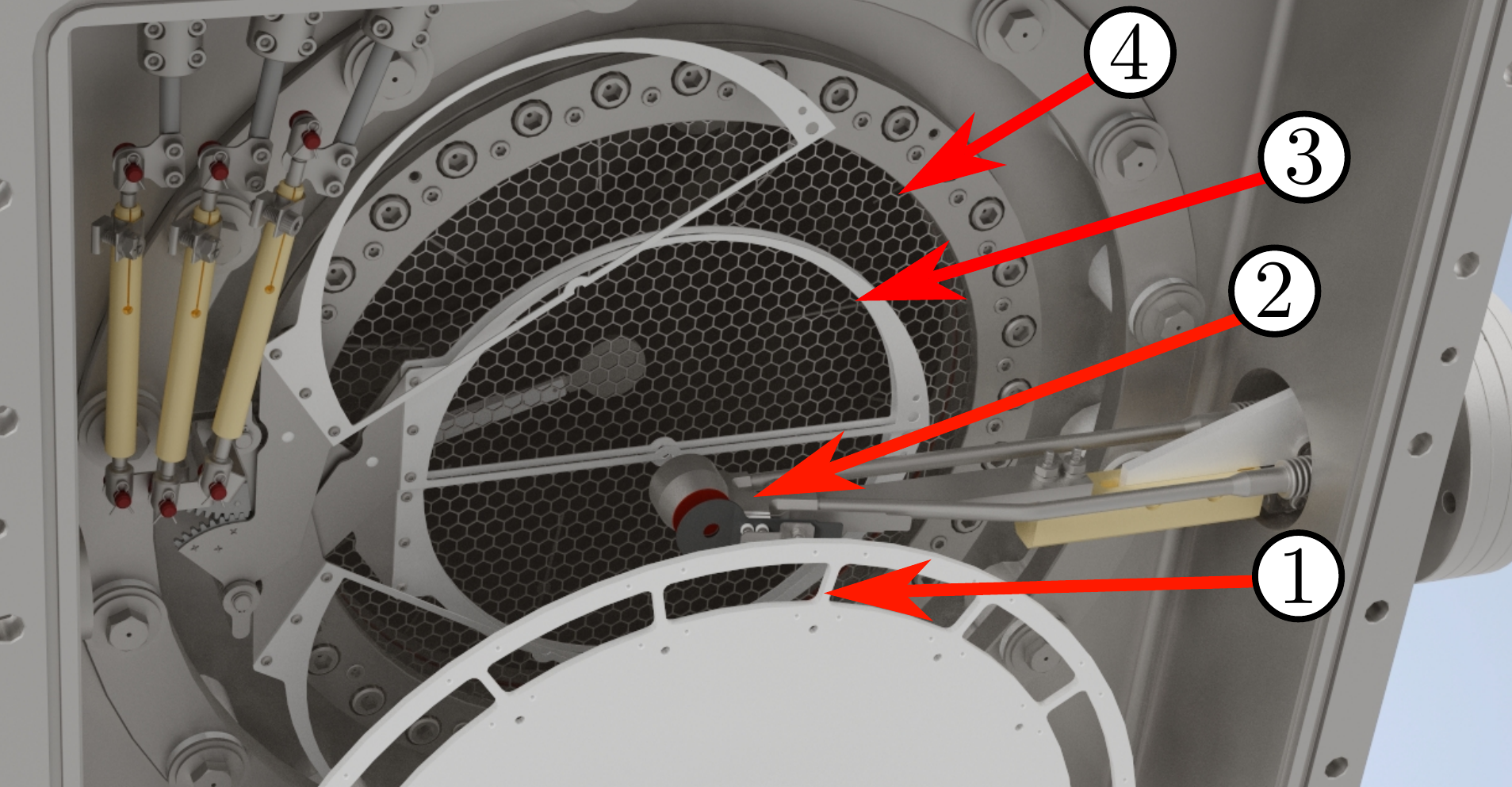}
\caption{CAD Rendering of the contents of the target box showing (1) the target wheel, (2) the beam dump, (3) the Pac-Men degraders (both open and closed) and (4) fixed degrader and window. \label{fig:TargetBox}}
\end{figure}

Rare isotope beam production at the N=126 Factory begins in the so-called ``Target Box'', seen in Fig.\,\ref{fig:TargetBox}.  Here, the 7-10 MeV/u stable primary beams from ATLAS are impinged on the target wheel, containing fifteen individual targets and one blank target position to aid with beam tuning. To prevent sublimation of the targets, the target wheel must be spun at hundreds of RPM, which is accomplished using an Applied Motion HT23-598 stepper motor, and also utilizes a Same Sky differential encoder to send a veto signal to a beam sweeper located at the beginning of ATLAS to stop beam from hitting the spokes of the wheel. Currently the reaction used is of $^{136}$Xe on 5 mg/cm$^2$ targets of natural-abundance platinum, which has been studied for production at the $N=126$ shell closure, with the promising calculated cross-sections \cite{Karpov17,Li19} validated for yields closer to stability \cite{Watanabe13}. 

Immediately after the target wheel there is a high-power beam dump covering from 0 to $5^\circ$ forward of the reaction point \cite{Savard20}. This is effectively a water-cooled Faraday cup, which consists, traveling along the path of the beam, of a collimator ring, electron-suppression electrode, beam dump, and backplate, and serves to stop unreacted and forward-scattered primary beam to avoid saturating the gas catcher while allowing for the tuning of beam on-target.

The MNT reaction products in the cone between 5 and $60^\circ$ then must be slowed before they enter the gas catcher and are stopped, which is accomplished via an array of degraders. These include a set of removable ``Pac-Man'' degraders consisting of three pairs of half-circles that are hinged at one side, allowing them to be opened and closed using linear pneumatic actuators and thus be removed or inserted from the path of the reaction products independently. Currently, the three pairs of Pac-Men degraders have 
one, two, and four 195 $\upmu$g/cm$^2$ layers of aluminized mylar installed, allowing for one to seven layers of 195 $\upmu$g/cm$^2$ to be inserted at any time. There is also a fixed degrader, which is installed directly on the frame of the aluminum window and has a radially-variable-thickness of between one and six 1.71 mg/cm$^2$ aluminum foils attached to an aluminum mounting ring. The thickness of the aluminum foil for a given angle is determined based on the predicted angular distribution of reaction product energies from GRAZING \cite{Winther94,Winther95} and stopping calculated with SRIM \cite{Ziegler2010}. Once slowed by these degraders, the MNT reaction products then go through the 3.3 mg/cm$^2$ aluminum window and into the gas catcher. 

\subsection{Gas Catcher and 90-Degree RFQ}
Though slowed by the degrader system, the MNT products are still moving at angles from 0$^\circ$ to 60$^\circ$ with respect to the beam axis, and need to be radially confined to produce a usable beam. The ``catching'' of the reaction products is performed using a large-volume RF gas catcher similar to that used in the CARIBU/nuCARIBU facility \cite{Savard16,McLain2022}. Geometric details of the gas catcher can be found in Ref.\,\cite{Savard20}. The gas catcher is filled with tens of mbar of high-purity He gas, and surrounded by a large outer box kept at low vacuum to ensure outward leakage of He and maintain purity. Collisions with the He gas cool the MNT products and result in ions in either a $1+$ or $2+$ charge state, depending on the second ionization potential. These ions are then confined radially using RF fields, and directed towards the downstream nozzle of the gas catcher using a combination of He flow and DC fields on the order of hundreds of volts. These ions are then extracted as a collimated beam through the nozzle. The stopping and extraction process occurs on the order of tens of ms, and is chemically-independently for all the products of an MNT reaction.

After extraction from the gas catcher, the rare isotope beam is bent 90$^\circ$, such that the downstream diagnostics are not in the path of the neutrons produced in the MNT reaction, and the helium gas load from the gas catcher is pumped out. This is accomplished in a bent radiofrequency quadrupole (RFQ) structure, which confines the ions radially while a DC drag field of tens of volts is applied to extract the beam. The RFQ is divided into three sections by PEEK baffles surrounding the electrodes and alumina rods between them, forming a differential pumping aperture. The first and second sections of the RFQ are pumped by separate Osaka TMP440 Turbomolecular pumps backed by either a Leybold Screwline SP630 dry screw vacuum pump or an Ebara A30W multistage dry vacuum pump, while the third section is pumped on by a Leybold Turbovac 350i Turbomolecular pump backed by a Leybold Ecodry 40 plus dry-compression multistage Roots vacuum pump. Together, these reduce the pressure in beamline after the RFQ to the order of $10^{-6}$ mbar.
\subsection{Separation and Preparation}
The beam which exits the 90$^\circ$ RFQ is a cocktail beam of the various isotopes and charge states that emerge from the gas catcher. Preliminary separation is achieved by first accelerating the beam across a 15-20 kV potential difference and directing it through a bending magnet of $R\sim10^3$. The magnetic field strength can be tuned as to bend the charge-to-mass ratio of interest into the center of the beamline, and mechanical slits are used to block the neighboring isobars. The resulting isobarically pure beam is then slowed back to a few hundred volts.

The beam is then injected into the RFQ Cooler-Buncher which is used to reduce the emittance of the isobarically separated beam from the separating magnet and bunch it in preparation for final separation in a multireflection time-of-flight mass separator. This RFQ Cooler-Buncher, described in detail in Ref. \cite{Valverde19b}, uses the design developed for FRIB's EBIT \cite{Lapierre18}. Its design features separated cooling and bunching sections to allow for a high-pressure region for efficient cooling but a lower-pressure bunching region to reduce collision-induced reheating, as well as injection optics optimized to increase acceptance. Additional copies of this buncher design are also in use at the St. Benedict experiment \cite{Valverde19a,Burdette26} and ATLAS's stopped beam area for low-energy CARIBU experiments \cite{Maass25}.  

After exiting the RFQ Cooler-Buncher and progressing through four 45$^\circ$ electrostatic deflectors, ion bunches enter the Mass Separation through Gated Reflection Time-of-Flight device (MSGR-TOF), the Notre Dame multi-reflection time-of-flight mass separator \cite{SCHULTZ16}. Ions are reflected between two electrostatic mirrors each consisting of five electrodes for a defined number of turns to allow separation in time-of-flight relative to each isotope's unique mass. Ions are then ejected from MSGR-TOF and pass through a Bradbury-Nielsen Gate \cite{Brunner12b} for selection of a unique time-of-flight range to clean away unwanted isobaric species. MSGR-TOF was designed and commissioned offline at the University of Notre Dame before installation at the N=126 Factory \cite{Liu21}. 

Diagnostic stations, consisting of an phosphor-screen MCP with a mirror (Beam Imaging Solutions BOS-18-IW) and a silicon detector (Ortec TB-020-300-1000), are located between all major components and ion-optical elements downstream the high-pressure region. Immediately following the 90$^\circ$ RFQ, a Faraday Cup is also included in the diagnostic station.

\subsection{Experimental Endstations} \label{sec: endstations}
The three initial experimental endstations for the N=126 Factory will focus on low-energy measurements of nuclear properties. The Canadian Penning Trap (CPT) mass spectrometer has been used for high-precision mass measurements at ATLAS for a quarter-century \cite{Savard01}. The phase-imaging ion-cyclotron-resonance technique \cite{Eliseev13} was implemented during the CPT's campaigns at ATLAS's CARIBU facility \cite{Orford20,Ray25}, which significantly increases the achievable precision and sensitivity of the CPT over the previous technique. The CPT is currently being reassembled after moving from CARIBU.  

The Reconfigurable Array for Coincidences and Cascade-Oriented Observational Nuclear Spectroscopy (RACCOONS) is currently under development at ANL. The RACCOONS decay station will incorporate the existing SATURN $\beta$-scintillator detectors and tape station \cite{XArray_SATURN}, the conversion-electron spectrometer LACES \cite{LACES}, and up to nine germanium detectors. This allows for $\beta$-decay half life measurements and other nuclear structure studies using beams from the N=126 Factory.

The Precision Optical Spectroscopy Experiment of Ions and atoms from Darmstadt Online at the N=126 (POSEIDON) is currently under development, with aims to mirror the ATLANTIS laser spectroscopy beamline at CARIBU \cite{Maass25}. This laser spectroscopy facility will enable nuclear structure studies of neutron-rich isotopes from the N=126 Factory, primarily through measurement of nuclear charge radii and electromagnetic moments. Commissioning of beam preparation elements, including an additional Cooler-Buncher, is ongoing.
\section{Status and Conclusion} \label{sec: status}
\begin{figure}
\includegraphics[width=\columnwidth]{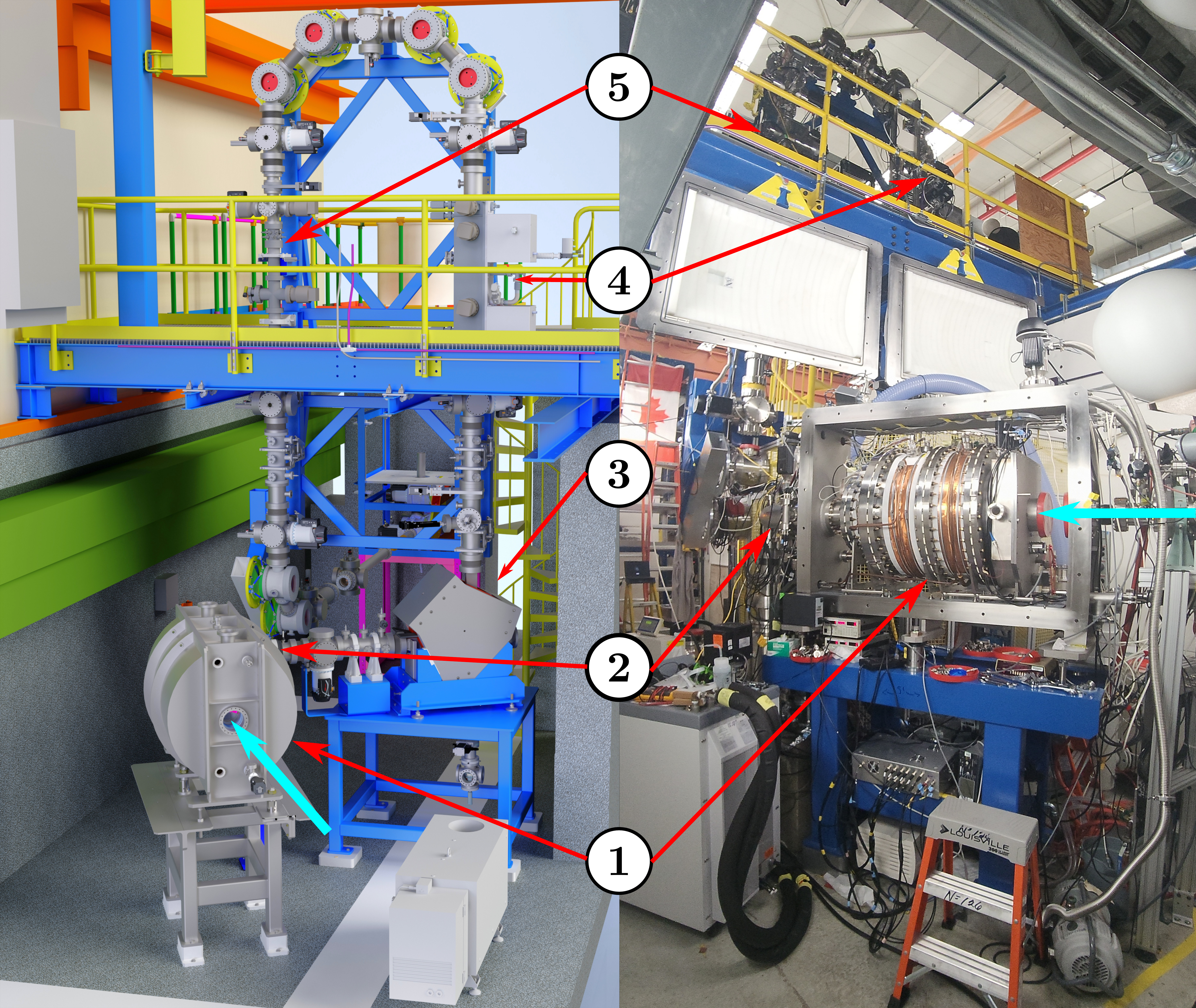}
\caption{(Left) CAD rendering of the finished N=126 Factory, shown facing northeast from the southwest wall, and (Right) Photograph of the current N=126 Factory, with the gas catcher box doors removed to show the gas catcher, taken facing northwest from the western corner of Area 126. Markers indicate (1) the Gas Catcher, (2) the 90$^\circ$ RFQ, (3) the magnetic separator, (4) Cooler-Buncher, and (5) MSGR-TOF, with cyan arrows indicating the path of the primary beam into the gas catcher.}\label{fig:Installation}
\end{figure}

Physical assembly of all the N=126 Factory components has been completed, and all have been installed in Area 126 (see Fig. \ref{fig:Installation}). 
Commissioning is ongoing, with $\sim$8 MeV/u $^{136}$Xe beam at up to $\sim$80 pnA (of the planned 5-10 p$\upmu$A \cite{Savard20}) being delivered by ATLAS. Several MNT products have been produced, stopped, and identified using the $A/q$ selectivity of the bending magnet and half-life measurements of implanted beam on the Si diagnostic detectors. Further, a californium source and an alkali thermal ion source are being used for offline commissioning of devices downstream of the gas catcher, and experimental stations are also being prepared.

Nuclear data for neutron-rich heavy isotopes is of considerable interest throughout the chart of the nuclides, and properties of nuclei near the $N=126$ shell closure are of particular interest for the study of the heaviest r-process peak. The measurement of properties of these nuclei is impeded by the difficulty of their production. The N=126 Factory at ANL's ATLAS User Facility will make use of MNT reactions to produce these isotopes, enabling their study with a suite of low-energy experimental apparatus. 

\section*{Acknowledgments}
This work is supported in part by the U.S. Department of Energy, Office of Nuclear Physics, under Contract No. DE-AC02-06CH11357; by the National Science Foundation under Grant No. PHY-2310059; by the Deutsche Forschungsgemeinschaft (DFG, German Research Foundation) -- Project-ID 279384907 -- SFB 1245; by the University of Notre Dame; and with resources of ANL’s ATLAS facility, an Office of Science User Facility.
\bibliographystyle{model1a-num-names.bst}
\bibliography{EMIS}
\end{document}